# Software Must Move!
## A Description of the Software Assembly Line

Martin J McGowan III
William L. Anderson

Product Assurance Department
Computer Consoles, Inc
97 Humboldt Street
Rochester, New York 14609

ABSTRACT

This paper describes a set of tools for automating and controlling the development and maintenance of software systems. The mental model is a software assembly line. Program design and construction take place at individual programmer workstations. Integration of individual software components takes place at subsequent stations on the assembly line. Software is moved automatically along the assembly line toward final packaging. Software under construction or maintenance is divided into packages. Each package of software is composed of a recipe and ingredients. Some new terms are introduced to describe the ingredients. The recipe specifies how ingredients are transformed into products. The benefits of the Software Assembly Line for development, maintenance, and management of large-scale computer systems are explained.

"*We shall never be wholly civilized until we remove the treadmill from the daily job.*" — Henry Ford

Introduction

What is the most expensive part of software development? It is not coding — coding typically accounts for only one-sixth of software's total cost. Instead, the real expenses are incurred during integration and testing. The alarming truth is that three times the coding costs — fully one-half of the money actually spent — is consumed by putting all the pieces of a large system together and making sure that they work.[l]

Why are these things which are so clear-cut in theory so expensive in practice? First, look at the facts. Integration and test procedures are tedious and repetitious. Large systems always have many components, and there are even more interconnections among these components. For a human, keeping track of such complexity is difficult. Mistakes are thus commonplace. Pieces are left out, or put together in the wrong order; proper connections among all of the contributing pieces are a rarity. As a consequence, one never knows when the job is finished. Clearly, then, the reason for the disproportionate costs of software integration and testing is that they are processes running "out of control!"

What is the solution? To begin with**,** people must stop doing the kinds of jobs that ought to be done by machines. Specifically, three things must be done:

1. break these large, unwieldy tasks into many smaller, manageable parts;
2. automate the procedures; and
3. move the software after each integration step.

The solution is to build the final system in stages. Gather pieces together incrementally. At the same time, test at each increment. Incremental testing provides measurable milestones. It also serves to certify the pieces. Once individual pieces are certified, certification of the whole system increases with each successive integration. Hence, managers and programmers gain confidence together — in both the package, and their ability to build and maintain it. The key is to keep the complexity of each integrated package small enough to be controlled.

It is true, that by increasing the number of integration steps one also increases the number of interconnections. But this increased complexity can be strictly and automatically controlled. Controlled complexity among the packages in a system is certainly an acceptable price to pay for certification of the final product.

As for automating the repetitious activities, there is no mystery involved. All that is needed is to write down the rules for each integration and testing action. Then these rules can be put into procedures, and can be executed by a computer. Although this may seem time consuming at first, it is an investment fully justified by its long-term benefits. Most large computer systems are maintained for many years. It is likely that during their lifetime testing and integration procedures will be repeated many times. Thus, it pays to put in the time and effort now — before it is needed — to automate these activities once and for all.

Why Software Must Move

After each integration step, the software <u>must</u> be moved to another place. Why? There are three major reasons.

1. Maintenance of a package cannot depend on hidden information. Software must be removed from the programmer's private domain.
2. Software must be isolated for testing. This applies to individual modules as well as the largest integrated piece.
3. Software to be delivered as a product must be distinguished from the tools used in building and testing.

First, moving software away from the programmer's private workspace exposes any hidden dependencies. For example, a program could depend on a macro definition file that is not available to everyone. Or the programmer may have built some special tools that are required for construction (and hence maintenance) of certain programs. These dependencies are quickly uncovered when software is moved to another place. Once exposed, hidden dependencies become explicit dependencies, acknowledged as part of the product.

Second, testing is facilitated in a controlled environment. Enforcing control is easier when the test environment is physically separate from the development environment. It follows that testing requires movement of software from development places to testing places. In addition to individual pieces, each integrated unit must be moved for testing. Eventually, the entire product must be moved for a full system test.

Finally, it is crucial to distinguish tools from products. Integration of components depends on the availability of tools. Testing depends on tools and test data. But the final product does not depend on these tools or data when it is delivered. The most effective way of cleaning up a software product is moving it to a clean location.

<u>Operation is Different from Development</u>

In 1977 Evan Ivie published a paper entitled "The Programmer's Workbench — A Machine for Software Development."[2] Ivie recognized that the activities of software development and maintenance are <u>fundamentally different</u> from the activities of operating a finished system. He proposed dedicating a separate computer to development and maintenance activities. The programmer's workbench machine could then be designed specifically for developing and maintaining computer software and its documentation.

There are several advantages to the workbench idea. In Ivie's own words:

> *Such an approach would help focus attention on the need for adequate tools and procedures; it would serve as a mechanism for integrating tools into a coordinated set; and it would tend to add stability to the programming environment by separating the tools from the product (the current approach is equivalent to carpenters leaving their tools in each house they build)*

The Programmer Workbench idea of separate development and target machines implies movement of software. A full system test for any large project requires software to be moved to a target machine. In fact, all embedded software is built on one machine and executed on another. Portable software is designed for movement to many machines. From these typical instances, it is clear that integration and testing of software always requires movement.

An important property of movement is its visibility. Movement of software can be strictly controlled by management, and the most efficient way to move software is automatically. Automated movements can be monitored — they provide measurable milestones.

<u>What Is the Software Assembly Line?</u>

The Software Assembly Line is a compact solution to the integration problems of large-scale system construction. It provides tools for building individual software components, and the environment for integrating these components into larger units. It actively promotes the process of dividing the system into small functional pieces. It then controls the building of each piece, and automates the integration procedures.

```
system = 1 { package } N
package = recipe
    + 0 { ingredient } M
    + 1 { output } K
```

Ingredients fall into one of three classes:

| Class | Description |
|---|---|
| <u>input</u> | An external ingredient that is transformed into output(s). |
| <u>primary</u> | A local ingredient that is transformed into output(s). |
| <u>tool</u> | An external ingredient that is used in building output(s). Tools are not transformed during the build process. |

Table 1.
Software Assembly Line Notation

In order to describe the Software Assembly Line further it is necessary to define some new terms. Individual components of systems are called packages. A package is composed of a recipe and ingredients. Ingredients are

  

further divided into primary elements, input elements and tools. These definitions are summarized in Table 1.

In this paper, extensive use is made of dataflow diagrams and data dictionary notation. For a full description see Tom DeMarco, <u>Structured Analysis and System Specification.</u> [3] In DeMarco's notation, curly braces "{" and "}" represent iteration, hence the notation in Table 1 is read as follows. A system is composed of at least one, and as many as N, packages. A package is composed of a recipe, zero or more (up to M) ingredients, and one or more (up to K) outputs.

An example of a simple package is shown as a dataflow diagram in Figure 1. It represents a package that compiles the files **iodefs** and **mainA** into an executable program, **programA.** Here, the file **iodefs** is an input element that contains input/output data structures. Input elements are ingredients that are <u>external</u> to a package. They are often shared by two or more packages.

Now look at the files **mainA** and **recipeA.** They are primary elements. Primary elements are those ingredients that are <u>local</u> to the package. **MainA** is the source code for **programA. RecipeA** is the procedure that is used to build **programA.**

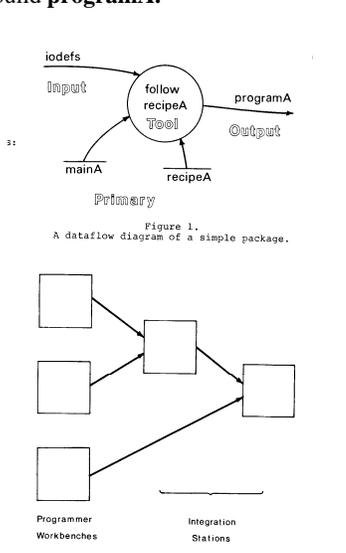

Figure 1.
A dataflow diagram of a simple package.

Finally, the follow procedure is a tool. Tools are ingredients used to transform the inputs and primaries into outputs, or desired products. Tools are not changed during the transformation. The file **programA** is the output of the package shown in Figure 1.

A Software Assembly Line is pictured in Figure 2. Here, each square represents a station on the assembly line. Arrows represent deliveries of software packages from one station to another.

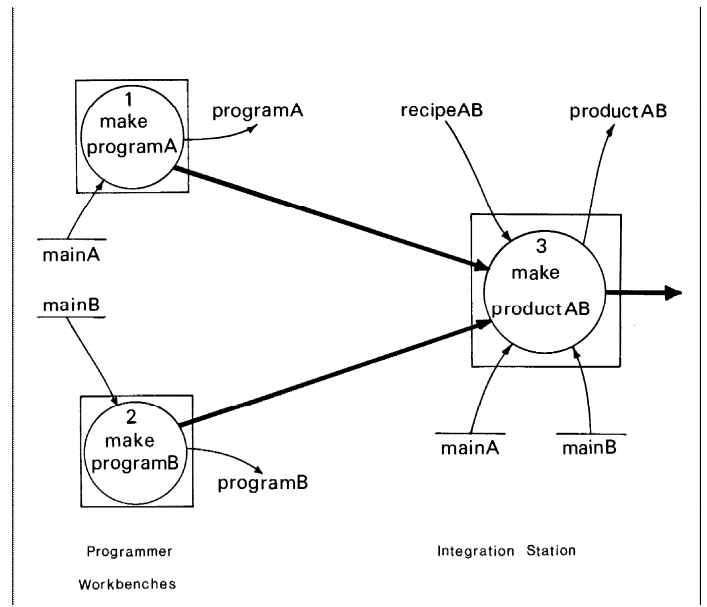

Figure 3.
Integration of two packages on the Software Assembly Line. Heavy arrows represent deliveries

The individual packages are built and tested at the first level stations, the programmer workbenches. Integration is therefore the process of combining individual packages into larger packages. This results in moving the primary elements of packages one step along the assembly line. Each subsequent station reflects yet another step in the integration process. Finally, at the last station, the output is a deliverable product.

But how does the Software Assembly Line work? Look at Figure 3. It shows the integration of Package 1 with Package 2 to form a larger unit, Package 3. In this example, Package 1 makes **programA** from the primary file **mainA,** and Package2 makes **programB** from the primary file **mainB.** Programmers at their workbenches have written the text of the primary elements of each package. The outputs, **programA** and **programB,** are tested and certified independently at the programmer workbenches.

Once certified, the packages are ready for integration. The primary files, **mainA** and **mainB,** are physically copied to the integration station where they form the integrated unit, Package 3. This copying procedure is called a delivery. Package 3 has two primary elements: **mainA** and **mainB. RecipeAB** provides the instructions to combine **productA** and **productB** into the more useful output, **productAB.** The distinction between **productA** and **productB** is not lost in the combination. **RecipeAB** has already been built and tested at some other workbench that is not shown in Figure 1. **ProductAB** is then tested and certified at this integration station. When testing of

Figure 2.
The Software Assembly Line.
Squares represent assembly stations;
arrows represent deliveries.

**productAB** is complete, the integrated package is ready for delivery to the next assembly station

Software Exists in Different States

In 1980 Paul Cashman and Anatol Molt described a software maintenance system based on communication among responsible parties.[4] The MONSTR system reveals three characteristics of software maintenance activities that are applicable to development as well. The first is simply that software exists in different states. For example, packages can be released, under development or repair, certified by integration or testing, etc. The second characteristic is that responsibility for states, and for the software in each state, is divided among the groups and individuals of an organization. Finally, the third characteristic is that transitions between states provide measurable milestones of progress. This is the value of a state model — states and transitions between them are well defined.

The Software Assembly Line, as shown in Figures 2 and 3, is an example of a state diagram. Packages at programmer workbenches are in the development or repair state. Each integration station might represent a higher degree of certification.

An individual, or group, is assigned responsibility for the activities at each assembly station. The responsible party owns that particular station. Deliveries are controlled by communication between the owners of the current and subsequent stations. When the package is certified at the current station, the owners of the next station execute the delivery procedure. After certification at the new station, responsibility is transferred to the owners of the new station. The discipline with which tools are employed at a particular station is the responsibility of the owner.

The Software Assembly Line automates deliveries. One attribute contributing to management and programmer confidence in system software is the ability of packages to withstand frequent deliveries. Frequent deliveries between assembly stations facilitate the integration activity and provide management information on the progress of the project. The number of assembly stations and the frequency of deliveries are planned to suit the integration requirements of the project.

Integration Assembles Packages, not Modules

Building large computer systems is the activity of bringing many distinct modules together into a single package. This integration activity is largely tedious and repetitious. For example, individual modules must be compiled and loaded, using particular libraries in a prescribed order. Compilation may be preceded or accompanied by inclusion of global data structure definitions. The results of these processes are executable programs, object module libraries, system support utilities, etc. The output are often packaged with an operating system, or embedded in a larger system, to form a deliverable product.

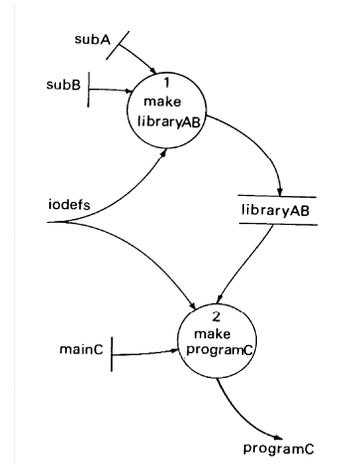

Figure 4.
The relations between two packages.

Most large systems have many interconnections among the constituent modules. The number of interconnections makes such systems difficult to build and maintain. Keeping track of coupling among modules is the main problem of software development. For example, Figure 4 shows the coupling between two simple packages. The final output is **programC.** Package 1 makes **libraryAB** from three files: **iodefs, subA,** and **subB.** Package 2 makes **programC** from **iodefs, mainC,** and **libraryAB.** Both pack-ages use **iodefs** as input. However, **libraryAB,** is an input to Package 2. It is clear that **libraryAB** must be built first — before **programC** can be made.

This simple example shows the kind of temporal dependence common among components in a large system.

In 1976, Frank DeRemer and Hans Kron argued for a module interconnection language capable of expressing the overall structure of a system.[5] DeRemer and Kron called the building of systems from distinct modules programming-in-the-large. They contrasted this with the building of individual modules: programming-in-the-small. In terms of the example in Figure 4, the dependency of **programC** on **libraryAB** is part of programming-in-the-large. The making of **libraryAB,** on the other hand, is an instance of programming-in-the-small. In the language of DeRemer and Kron, **libraryAB** is a "resource" used by the package that makes **programC.** In the language of the Software Assembly Line it is an input.

DeRemer and Kron list four objectives for a module interconnection language. Such a language should serve as:
    1. a project management tool encouraging incremental

- development and testing;
2. a design tool for establishing the overall system structure, including the integration of subsystems;
3. a communication tool permitting hierarchical structure and disciplined connections among packages; and
4. clear and formal documentation of the system structure.

Make and Other Tools

The UNIX[*] program make[6] and the Software Assembly Line satisfy these objectives for a module interconnection language. Make is a language that provides a syntax for specifying hierarchical dependencies. Hierarchical, or structured, partitioning is the usual way to build computer programs. Structured Design [7] was originally developed to help control the complexity of individual software packages — the problems of programming-in-the-small. It is commonly used to

1. build computer programs from source code;
2. build libraries of software modules;
3. integrate a document from its pieces; and
4. integrate and test previously built programs.

The Software Assembly Line, on the other hand, grew out of the need to control the complexity of systems — the problems of programming-in-the-large. The Software Assembly Line is a machine (an organized set of tools) that uses make as its defining language.

Make uses a description file called a makefile in which the programmer writes down the dependencies among the output, input, and primary files, and the exact sequence of operations needed to build a new version of the outputs. In terms of the Software Assembly Line, the makefile is the recipe for the package.

Briefly, a makefile entry has the following general form:
```
target: component1 ... componentN
     command line1
     command line2
         .
         .
         .
     command lineM
```

Component files are either inputs or primaries. Outputs are always targets. The tools, and how they are used, are specified in the command lines.

For example, a makefile for the package shown in Figure 1 might contain the following text:

```
programA: iodefs mainA
     cc mainA -o programA
```

The first line of this text specifies that **programA** depends on the files **iodefs** and **mainA.** The second line specifies the command needed to compile **mainA** into **programA**. (The "-o" instructs the C compiler, cc, to put its output into a file named **programA**.) If either **iodefs** or **mainA** (or both) are updated, make executes the command on the second line, building a new copy of **programA**.

The other tools that define the Software Assembly Line break down the ingredients of a package into tool, input, primary, and output elements. This list is called the tipo (pronounced "tip-oh") list and it may be generated whenever it is needed. For the example of Figure 1 the list looks like

```
TOOL = cc
INPUT = iodefs
PRIMARY = mainA
OUTPUT = programA
```

Karen Huff, in a paper describing configuration management in a programming environment, discusses the relations among the ingredients of a package.[8] In her view, "tool inputs and outputs are the configuration items of interest." However, Huff states that effective configuration management must be enforced in the programming environment. The Software Assembly Line provides a tipo list for each package. The dependency of a package on external files and the tools needed to build its outputs can be included in each makefile. This list of tools, inputs, primaries, and outputs is the information necessary for tracking and controlling connections among packages. The tipo lists for all packages of a system could be used to produce complete documentation of the overall system structure.

Huff goes on to say that it is not enough to know simply that a certain tool is required. To have confidence in a tool, one must also know its history. What version of the tool is available? When was it released? What are its certifications? These are important questions. Confidence in the output of a package is based on confidence in the ingredients. The Software Assembly Line offers a method for certifying all ingredients. Tools and inputs built and maintained on the assembly line can be tested and certified in the same way as other products. In fact, certification of products starts with certification of tools. Certified primary and input elements, transformed by certified tools, produce certified outputs.

Summary

The Software Assembly Line provides automated tools for two essential activities. One is building packages of software. The other is delivery of packages between stations on the assembly line. These two activities are sufficient to automate the integration of software systems.

Building products incrementally and automating

---

[*] UNIX is a Trademark of Bell Laboratories

the activities yield several benefits.

1. Costs are reduced — labor intensive tasks are shifted to the machine.

2. Control over the activities is increased. Automated tools provide measurable milestones.

3. Reliability of the final product is increased. Confidence in the procedures and tools grows with use.

At Computer Consoles, Inc., the Software Assembly Line is being used to maintain and certify the Assembly Line itself, existing Computer Consoles products, and the utilities of the UNIX operating system.

Ivie, in the conclusion of his paper on the Programmer's Workbench, remarked that the programming profession needed a methodology that could be "transferred from one project to another and from one machine to another."[9]  The Programmer's Workbench was a step in the right direction.  The Software Assembly Line is an extension of Ivie's basic ideas.  It is applicable to any project, and implementation is not restricted to any specific machine.  It is another step toward a general development and maintenance methodology.

Acknowledgements

The realization of the Software Assembly Line is due in part to the generous support of the management at Computer Consoles, Inc.  Mr. Lewis A. White, Director of Systems Engineering, and Mr. Eiji Miki, Director of Product Assurance, deserve special mention in this regard. In addition, many individuals at Computer Consoles, Inc., have helped form the current shape of the assembly line.  In particular, David P. Somers made significant improvements to the UNIX m4 macro processor, a major tool of the Software Assembly Line.  In addition, Allen Brumm and Michael L. Sortino made significant contributions to the current implementation.  Finally, Mr. David C. Smith, of I. P. Sharp Associates, Inc., deserves special thanks for his careful criticism and editorial help.